# Normal-state electrical resistivity and superconducting magnetic penetration depth in $Eu_{0.5}K_{0.5}Fe_2As_2$ Polycrystals

V.A. Gasparov[1], H.S. Jeevan[2] and P. Gegenwart[2]

[1] *Institute of Solid State Physics RAS, Chernogolovka, 142432, Russian Federation*
[2] *I. Physik. Institut, Georg-August-Universität Göttingen, D-37077 Göttingen, Germany*

We report measurements of the temperature dependence of the electrical resistivity, $\rho(T)$, and magnetic penetration depth, $\lambda(T)$, for polycrystalline samples of $Eu_{0.5}K_{0.5}Fe_2As_2$ with $T_c$=31 K. $\rho(T)$ follows a linear temperature dependence above $T_c$ and bends over to a weaker temperature dependence around 120 K. The magnetic penetration depth, determined by radio frequency technique displays an unusual minimum around 4 K which is associated with short-range ordering of localized $Eu^{2+}$ moments.

The recent discovery of superconductivity in LaOFeP at $T_c \approx$ 4K by Kamihara *et al.* [1] has lead to intensive studies on electron and hole doped iron arsenide oxide superconductors RFeAsFO (R=La, Sm) with $T_c$ as high as 55 K in SmFeAsO$_x$F$_{1-x}$ [2]. Very recently, Rotter *et al.* [3] found that the oxygen free iron arsenide $BaFe_2As_2$ in which Ba is partially substituted by potassium ions, is a superconductor below $T_c$=38 K, which was confirmed for $(K/Sr)Fe_2As_2$ compounds with $T_c$=37 K [4]. The FeAs layers common to both series of compounds seem to be responsible for superconductivity. Jeevan *et al.* recently observed that $EuFe_2As_2$ shows a spin-density wave (SDW) type transition at 190 K, and becomes superconductive below 32 K after partial substitution of Eu by 50%K [5]. Below about 10 K, short-range magnetic order of the $Eu^{2+}$ moments was suggested by a feature in the magnetic susceptibility. Here we focus at first on the temperature dependence of the normal-state resistivity and then on the superconducting magnetic penetration depth in order to probe the influence of local $Eu^{2+}$ moments on superconductivity.

Polycrystalline samples of $Eu_{0.5}K_{0.5}Fe_2As_2$ were synthesized from stoichiometric amounts of the starting elements Eu (99.99%), K (99.9%), Fe (99.9%), and As (99.999%) by solid-state reaction method under Argon atmosphere, as described in [5]. The sample crystallizes in the tetragonal structure with lattice parameters a=3.8671Å and c=13.091 Å [5]. X-ray analysis reveals that the composition of the samples is close to the expected 0.5:0.5:2:2 stoichiometry. Samples had form of rectangular bars of about 1.7×1.7×1.1 mm$^3$.

A standard four-probe *ac* (9Hz) technique was used for resistance measurements. A well-defined cubic geometry of the samples provided for the precise $\rho(T)$ measurements through van der Pauw four probe method and the superconducting magnetic penetration depth experiments. The temperature was measured with platinum (PT-103) and carbon glass (CGR-1-500) sensors. The measurements were performed in a liquid Helium variable temperature cryostat in the temperature range between 1.3 K and 300 K. Magnetic measurements of $\rho(T)$ and $\lambda(T,H)$ were carried out using a superconducting coil in applied fields of up to 3 T and at temperatures down to 1.3 K.

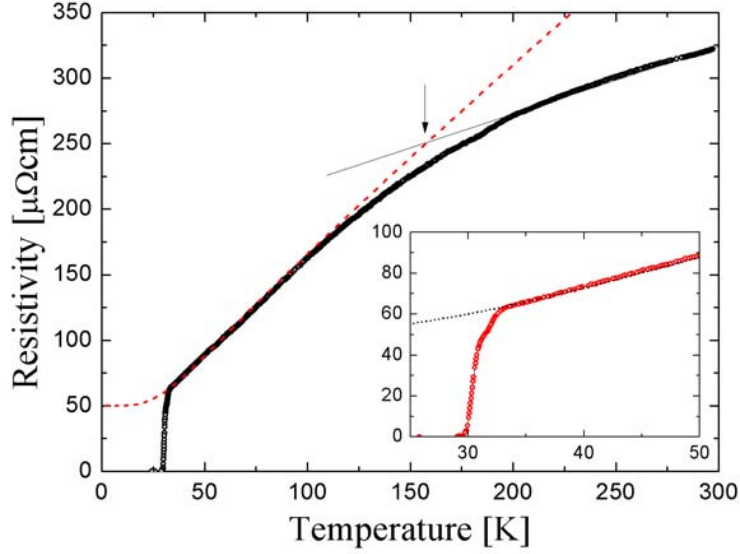

FIG. 1: (color online) Temperature variation of the resistivity of Eu$_{0.5}$K$_{0.5}$Fe$_2$As$_2$ sample. The inset shows the superconducting transition on an enlarged scale. Dashed line is a fit with B-G Eq.2 below 150K and solid line is extrapolation from ρ(T) above 150K.

We used a radio frequency LC technique [6] to measure λ(T) of Eu$_{0.5}$K$_{0.5}$Fe$_2$As$_2$ samples. This technique employs a simple rectangular solenoid coil into which the sample is placed. Changes in the magnetic penetration depth of the sample lead to the change of the coil's inductance L that in turn results in the change of the resonance frequency ω (2-20 MHz) of the LC circuit. The connection between parameters of the circuit and λ (T) is described by following simple equation [6]:

$$\lambda(T) - \lambda(0) = \delta \cdot \frac{\omega^{-2}(T) - \omega^{-2}(0)}{\omega^{-2}(T_n) - \omega^{-2}(0)} \tag{1}$$

Here δ=0.5√(c$^2$ρ/2πω) is the imaginary part of a skin depth above T$_c$, which was determined from the ρ(T) measurements [6], ω(T) is the resonance frequency of the circuit at arbitrary T, ω(T$_n$) and ω(0) are the same one's above T$_c$ and at zero temperature, respectively.

Fig.1 shows the normal-state resistivity ρ(T) of Eu$_{0.5}$K$_{0.5}$Fe$_2$As$_2$ sample at a doping $x$ = 0.5. Eu$_{0.5}$K$_{0.5}$Fe$_2$As$_2$ is a bad metal with a specific resistivity around 300 μΩ cm at room temperature. To emphasize the variation of ρ(T) in a superconducting state, we plot these data below 50 K in the inset. ρ(T) decreases smoothly with temperature, while drops abruptly to zero with a midpoint at T$_c$=31 K, which clearly indicates superconductivity. Above T$_c$, ρ(T) exhibits linear temperature dependence up to 120 K and develops a remarkably pronounced downturn from its linear-T behavior at higher temperatures. We first try to analyze the ρ(T) dependence in terms of the Bloch-Grüneisen (BG) equation for the electron-phonon (e-p) scattering:

$$\rho(t) - \rho(0) = 4\rho_1 \cdot t^5 \int_0^{1/t} \frac{x^5 e^x dx}{(e^x - 1)^2} \tag{2}$$

Here, $\rho(0)$ is the residual resistivity, $\rho_1 = d\rho(T)/dt$ is the slope of $\rho(T)$ at high $T > T_R$, $t = T/T_R$ and $T_R$ is the resistive Debye temperature. It is clear from Fig.1 that the BG model describes the $\rho(T)$ dependence below 120 K with rather low $T_R = 180$ K, suggesting an importance of the e-p interaction. However, we could not fit $\rho(T)$ in the entire temperature range with Eq.2 because the resistance bending over 120 K.

Such unusual $\rho(T)$ dependence in $Fe_2As_2$ compounds, is far from being clear and disputed in the scientific community. The abrupt changes in the $\rho(T)$ dependence at 150 K may be considered as a signature of a phase transition, where the crystal structure changes from tetragonal to orthorhombic, as was observed by Rotter et al. [7] at 140 K for different compositions of $Ba_{1-x}K_xFe_2As_2$. The reduction of the lattice symmetry was visible by (110)-reflections XRD peak splitting up to x = 0.2, however is absent for superconducting samples at x ≥ 0.3. Thus, the tetragonal to orthorhombic phase transition, as well as the magnetic (spin-density-wave) transition are completely suppressed in superconducting $Ba_{0.6}K_{0.4}Fe_2As_2$ [7]. At the same time the resistivity bending over at 120 K is still present [7].

Very recently, Gooch et al. [8] fitted the low-temperature part of $\rho(T)$ at T<100 K of $Ba_{1-x}K_xFe_2As_2$ to a power-law dependence, $\rho(T) - \rho(0) = AT^n$, and found evidence for quantum critical behavior: The exponent n sharply decreases with x from n=2 to n=1 near a critical concentration $x_c= 0.4$, and then increases again to a value close to 2 at x = 1 [8]. Furthermore, the thermoelectric power divided by temperature displays a logarithmic dependence $S(T)/T \propto \log T$ near critical doping. Both results would be compatible with a quantum critical point at $x_c$ which is hidden by superconductivity, similar as found in various heavy-fermion systems [9]. Whereas in the heavy-fermion case the characteristic magnetic energy scale is of the order of 10 K and quantum criticality is typically cut-off above this temperature, in $Fe_2As_2$ systems, the SDW transition takes place at about 200 K and thus, quantum criticality is expected to extend up to much higher temperatures. In this scenario, the observed crossover in $\rho(T)$ of $Eu_{0.5}K_{0.5}Fe_2As_2$ at 150 K would then mark the upper limit of the universal quantum critical regime in the system. Certainly, the existence of quantum critical fluctuations in $Fe_2As_2$ systems needs to be investigated by inelastic neutron diffraction or other magnetic probes. We also note, that the $\rho(T)$ dependence in $Ba(Fe_{0.93}Co_{0.07})_2As_2$ single crystals in the normal state remains almost linear up to room temperature [14].

We now turn to the magnetic penetration depth in the superconducting state of $Eu_{0.5}K_{0.5}Fe_2As_2$. Given that the $\lambda(T)$ dependence has a BCS form close to $T_c$:

$$\lambda(T) = \frac{\lambda(0)}{\sqrt{2 \cdot (1 - \frac{T}{T_c})}} \quad (3)$$

we plot $[\omega^{-2}(T) - \omega^{-2}(0)]/[\omega^{-2}(T_n) - \omega^{-2}(0)]$ data versus BCS reduced temperature: $1/\sqrt{(2(1-T/T_c))}$ in the region close to $T_c$. We use the slope of $\lambda(0)/\delta$ vs $1/\sqrt{(2(1-T/T_c))}$ and (3) to obtain an unusually large value of $\lambda(0) = 4.02 \times 10^{-4}$ cm from $\delta = 1.088 \times 10^{-2}$ cm. For a BCS-type superconductor with the conventional s-wave pairing form, the $\lambda(T)$ has an exponentially vanishing temperature dependence below $T_c/2$ (where $\Delta(T)$ is almost constant) [6]:

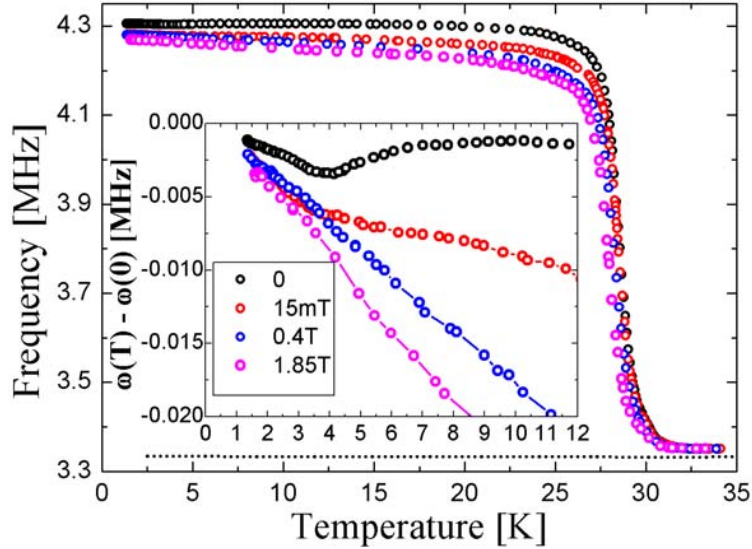

FIG.2. (color online) Temperature variations of resonance frequency of LC circuit ω(T) for Eu$_{0.5}$K$_{0.5}$Fe$_2$As$_2$ sample. The inset shows the temperature dependence of ω(T)- ω(0) in extended scale. The dashed curve is for empty coil.

$$\lambda(T) = \lambda(0) \cdot \sqrt{\frac{1}{\tanh(\frac{\Delta(0)}{2k_B T})}} \qquad (4)$$

for dirty limit, i.e. with short mean free path l<ξ [6]. Here Δ(0) is the energy gap.

In Fig.2 we compare the temperature dependencies of ω(T) behavior at rather small magnetic fields. As we can see from the inset, the low-T part of this dependence has unconventional minimum around 4 K, which become a break like in small magnetic field 15 mT, and completely disappear at larger field 0.4 T. Also, the magnetic field dependence of ω(T) is quite strong. On the other hand, the ω(T) curves clearly display a smooth variation below 3 K which simplifies the extrapolation of the resonance frequency ω(T) of our LC circuit down to zero temperature in order to calculate λ(T) from Eq.1. At the same time the existence of this minima makes impossible the exploration of the exponentially vanishing BCS temperature dependence according to Eq.4 below T$_c$/2 for the determination of Δ(0).

We plot in Fig.3 the deviation λ(H) - λ(T=0) as a function of the magnetic field at very small *H*. In contrast to measurements of the magnetic induction on PrFeAsO$_{1-x}$ [10], the λ(H) - λ(0) dependence displays a sharp signature in the magnetic field dependence with clear tendency towards saturation at 15 mT independently from temperature, while we expect a linear dependence with a break point at low fields caused by the Meissner effect [6]. The observed smooth minimum in λ(H) at 4.2 K has the same origin as ω(T) shown in Fig.2. This result indicates that there is no edge point in λ(H) close to the true field of flux penetration in striking contrast with magnetization data in PrFeAsO$_{1-x}$ used to deduce H$_{c1}$ [10]. Thus we could not determine the value of *H$_{c1}$* in contrast to e.g. the case of ZrB$_{12}$ [6], apparently due to possibly melting of the vortex solid and the presence of strong vortex pinning [14].

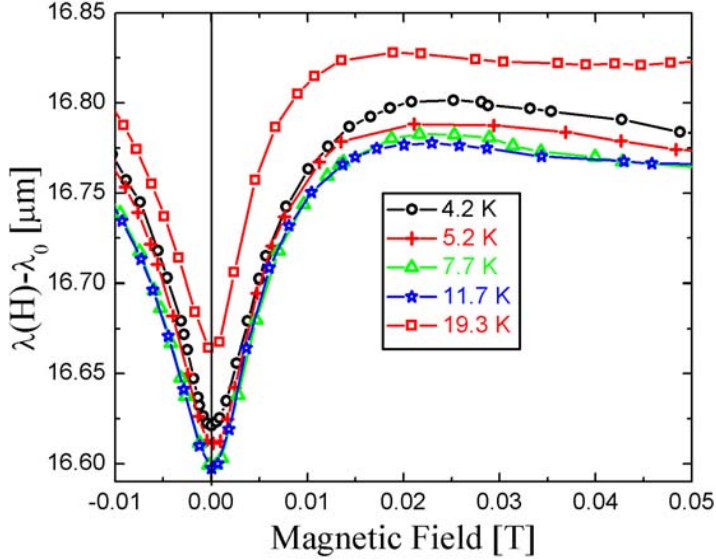

FIG.3. (color online) Typical magnetic field variation of $\lambda(H)$- $\lambda(0)$ of a Eu$_{0.5}$K$_{0.5}$Fe$_2$As$_2$ sample at different temperatures: 4.2 K, 5.2 K, 7.7 K, 11.7 K and 19.3 K. The solid lines are the guides for the eye.

In the absence of vortices we probe the London penetration depth $\lambda$. Important problems for $\lambda(T)$ measurements are: (i) the determination of the basic superconducting parameter $\lambda(0)$ and (ii) its temperature dependence, to see whether s-wave or d-wave pairing form exist. Both these problems can be addressed from the low-T $\lambda(T)$ dependence. However, one can easily notice from Fig.4 an unconventional behavior of the superfluid density $[\lambda(0)/\lambda(T)]^2$ of Eu$_{0.5}$K$_{0.5}$Fe$_2$As$_2$ at low temperatures. In contrast to BCS-type behavior, we observe a small but well defined anomaly with a pronounced minimum at 4 K. Small magnetic fields wash out this feature and strongly influence the superfluid density.

Apparently, the strong magnetic field dependence of $\lambda(T)$ is due to magnetic flux lines partially penetrating the sample in the vortex state of the superconductor. Very strong flux pinning was also observed by Eskildsen *et al.* [13] in Ba(Fe$_{0.93}$Co$_{0.07}$)$_2$As$_2$ single crystals with a disordered vortex arrangement. In our system the magnetic field will also affect the Eu ions. The observed anomaly in $\lambda(T)$ is very likely related to short-range ordering of the Eu$^{2+}$ moments coexisting with the superconducting state below 10 K, as seen in the magnetic susceptibility [5] and $^{151}$Eu Mössbauer spectroscopy [11].

The magnetic susceptibility anomaly at low-T was absent in (K/Sr)Fe$_2$As$_2$ compounds [3,4] as well as in the $\lambda(T)$ dependence for Ba(Fe$_{0.93}$Co$_{0.07}$)$_2$As$_2$ single crystals [12]. While the specific heat vs T signature associated with the superconducting transition provides clear evidence of the bulk nature of superconductivity in Eu$_{0.5}$K$_{0.5}$Fe$_2$As$_2$ [5], the rather large $\lambda(0)$ indicates an unusually large penetration of the electromagnetic field in this compound with composition close to the quantum critical point. We would like to stress that $\lambda(0)$ was determined from the temperature dependence of $\lambda(T)$ close to T$_c$ by assuming a BCS-like form, but not from low-T data, which are masked by magnetism of $Eu^{2+}$ ions. The influence of the short-range Eu-ordering on the lower-critical field and on the pinning behavior in $Eu_{0.5}K_{0.5}Fe_2As_2$ should be studied in more detail.

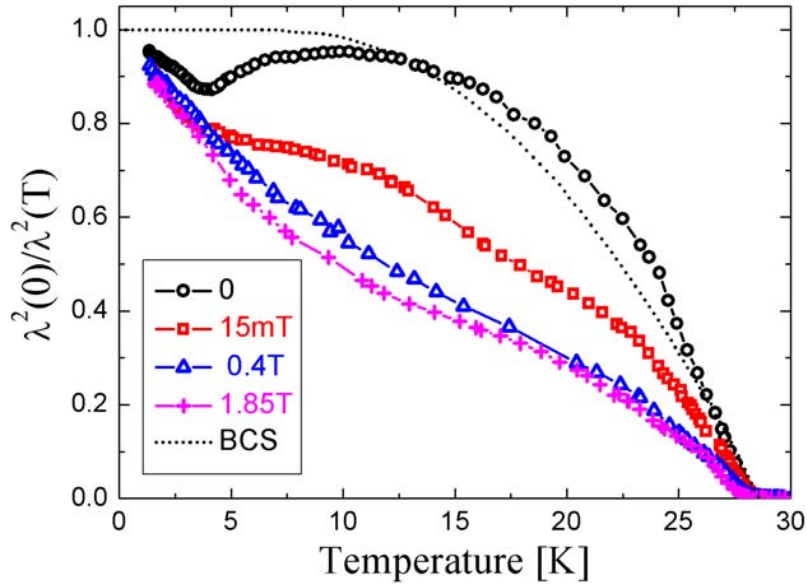

FIG. 4. (color online) Superfluid density, $[\lambda(0)/\lambda(T)]^2$, of the $Eu_{0.5}K_{0.5}Fe_2As_2$ sample in different magnetic fields for the $\lambda(0)=4.02*10^3$ nm. The predicted behavior of $[\lambda(0)/\lambda(T)]^2$ within the BCS model is shown by dotted line.

*In summary*, we have performed a systematic study of the temperature and magnetic field dependence of the resistivity, $\rho(T)$, and the magnetic penetration depth, $\lambda(T)$, on polycrystalline samples of $Eu_{0.5}K_{0.5}Fe_2As_2$. The $\rho(T)$ dependence may be described by the Bloch-Grüneisen formula only in a limited temperature regime below 120 K and bends over at higher temperatures. Alternatively, the observed $\Delta\rho \propto T$ dependence may be interpreted in terms of quantum critical behavior which is cut-off above 120 K. The superfluid density does not exhibit BCS type dependence and has an unconventional minimum close to 4 K, very likely due to a short-range ordering of $Eu^{2+}$ ions. Small magnetic fields destroy this signature. Altogether, our results indicate unusual normal and superconducting properties in $Eu_{0.5}K_{0.5}Fe_2As_2$ .

We would like to thank V.F. Gantmakher for very useful discussions. We acknowledge financial support by the RAS Program: New Materials and Structures (Grant 4.13), the DFG Research Unit 960 and BRNS (Grant No. 2007/37/28).